\documentstyle[amsfonts,preprint,aps]{revtex}

\begin{document}
\title{Do Fresnel coefficients exist ?}
\author{D. Felbacq$^{1}$,B. Guizal$^{1}$, F. Zolla$^{2}$}
\address{$^{1}$LASMEA UMR-CNRS 6602\\
Complexe des C\'{e}zeaux\\
63177 Aubi\`{e}re Cedex, France\\
$^{2}$Institut Fresnel UMR-CNRS 6133\\
Facult\'{e} des Sciences de Saint-J\'{e}r\^{o}me\\
13397 Marseille Cedex, France}
\maketitle

\begin{abstract}
The starting point of the article is the puzzling fact that one cannot
recover the Fresnel coefficients by letting tend the width of a slab to
infinity. Without using the so-called limiting absorption principle, we show
by a convenient limit analysis that it is possible to define rigorously the
field diffracted by a semi-infinite periodic medium.
\end{abstract}

\tightenlines

\section{Introduction}

Over the past decades, the numerical resolution of Maxwell equations has
made outstanding advances, thanks to the various rigourous methods developed
to solve them. One may think for instance, in the theory of gratings, of the
integral method, the differential method or the method of fictitious
sources. The tremendous progress of computer capacities has been equally
important. Generally speaking, it seems that we now have very efficient
tools to solve any diffraction problem (the 3D is merely a problem of
computer power, and not a theoretical one). Nevertheless, quoting R. Thom,
the famous mathematician, Field medalist and inventor of Catastrophe Theory,
''predicting is not explaining''. And although we are able to compute almost
everything, we cannot explain much. This paper points out that a simple
problem, here a one dimensional one, may lead to deep questions and somewhat
interesting answers. Our starting point is a basic remark concerning
diopters. The diffraction of a plane wave, in normal incidence, by a diopter
separating the vacuum from a homogeneous medium of optical index $\nu $
gives rise to Fresnel coefficients as is well known. 
\begin{equation}
r=\frac{1-\nu }{1+\nu },t=\frac{2}{1+\nu }
\end{equation}

This is indeed a fundamental result, commonly used by the opticists.
Unfortunately diopters do not exist and, consequently, the famous Fresnel
coefficients should not exist either. The problem may be formulated in the
following way: let us consider a slab of index $\nu $ and width $h$. This
gives rise to a reflection coefficient $r_{h}$. A natural question is: do we
get $r$ when letting $h$ tend to infinity in $r_{h}$? When we consider a
lossless material ($\nu $ is therefore real), the answer is no, obviously.
So, how can one measure Fresnel coefficients as there is no such thing as a
semi-infinite medium in practice ? Two kinds of answers are generally put
forward. The first answer that seems to come naturally to many is to evoke
the absorption, and the fact that transparent media do not exist. Let us say
that it is a poor answer. Indeed, consider for instance the diffraction of a
beam, in optical range, by a homogeneous layer (for example, silice for
which the refractive index is $\nu =1.47+i10^{-5}$), let's say of a few
millimeters (namely the substrate). We can hardly say that there is any
absorption in that case. But we have here a medium that is almost thousand
wavelengths wide. Another way out generally put forward is to evoque
coherence length. This argument is probably correct for the natural light
for which coherence length is about micrometer, i.e. less than the depth of
materials commonly used in optics (See \cite{BornandWolf} for an exhaustive
review on this subject). However for a beam whose coherence length is
greater than a few millimeters one has to find another explanation (it is
the case for He-Ne laser beam, for which coherence length is about 20 cm).
Without using the two above quoted arguments, we propose a rigorous answer
to this problem, in the more general case of a semi-infinite periodic
medium. More precisely, we show that it is possible to define the field
reflected by a semi-infinite periodic medium. We demonstrate that there is
no need to use the so-called limiting absorption principle nor to use
explicitly the notion of coherence length. We proceed in two stages. First,
we study the diffraction of a plane wave by a finite medium. In this case,
the diffracted field is characterized by a reflection coefficient. In a
second stage, we study the limit behavior of the reflection coefficient as
the width of the medium tends to infinity. All the fields that we consider
are $z$-independent and therefore we may reduce the diffraction problem to
the basic cases of polarization: E// (electric field parallel to $z)$ and
H// (magnetic field parallel to $z$).

\section{Reflection by a finite one-dimensional medium}

We consider a one dimensional structure made of $N$ identical layers (See 
\cite{Brillouin}). A layer is characterized by its relative permittivity $%
\varepsilon \left( x\right) $, which is assumed to be real and positive. For
convenience up to the section 4 included, the thickness of a layer is equal
to $1$. The structure is illuminated by a wavepacket of the following form: 
\begin{equation}
U^{i}(x,y,t)=\int_{{\bf P}}A(k,\theta )\exp (ik(\cos \theta x+\sin \theta
y))\exp (-i\omega t)d\mu \; ,  \label{incident}
\end{equation}
where $\mu $ is some measure over the set ${\bf P}$ of parameters $\left(
k,\theta \right) $: 
\begin{equation}
{\bf P=}\left\{ \left( k,\theta \right) \in {\Bbb {R}^{+}\times \left] -%
\frac{\pi}{2},\frac{\pi}{2}\right[ }\right\} \; .
\end{equation}

Explicit forms of $\mu $ will be given in section 4. By means of Fourier
analysis, the diffraction of such a wavepacket by a finite structure may be
reduced to the study of the diffraction of a plane wave, which we develop in
the following paragraph.

We consider a plane wave of wave vector ${\bf k}_{0}$ illuminating the
structure under the incidence $\theta $ ( ${\bf k}_{0}$ is assumed to belong
to the $(xOy)$ plane, we denote $\beta _{0}=k_{0}\cos \theta $). The total
field is described by a single variable function $U_{N}(x)$. When the
electric field is parallel to the z-axis (E// case) $U_{N}(x)\exp
(ik_{0}y\sin \theta )$ represents the $z$-component of the electric field
and when the magnetic field is parallel to the z-axis (H// case), it
represents the z-component of magnetic field . Denoting: $\beta \left(
x\right) =k_{0}\left[ \varepsilon (x)-\sin ^{2}(\theta )\right] ^{1/2},$ the
total field $U_{N}$ verifies the following equation:

\begin{equation}
0\leq x\leq N:\left( q^{-1}(x)U_{N}^{\prime }\right) ^{\prime
}+q^{-1}(x)\beta ^{2}U_{N}=0  \label{problem}
\end{equation}
and the radiation conditions lead to the following equations

\begin{equation}
\left\{ 
\begin{array}{l}
x\leq 0:U_{N}(x)=\exp (i\beta _{0}x)+r_{N}\left( k,\theta \right) \exp
(-i\beta _{0}x) \\ 
x\geq N:U_{N}(x)=t_{N}\left( k,\theta \right) \exp (i\beta _{0}x)
\end{array}
\right.  \label{propag}
\end{equation}
with: $q\equiv 1$ for E// polarization, $q\equiv \varepsilon $ for H//
polarization. Let $\chi_{1}$ and $\chi_{2}$ be the solutions of equation (%
\ref{problem})\ verifying

\[
\left\{ 
\begin{array}{c}
\chi_{1}\left( 0\right) =1,\;\chi_{1}^{\prime }(0)=0 \\ 
\chi_{2}\left( 0\right) =0,\;\chi_{2}^{\prime }(0)=1
\end{array}
\right. 
\]
The fundamental matrix of the system is then

\[
V(x):=\left( 
\begin{array}{cc}
\chi_{1}(x) & \chi_{2}(x) \\ 
\chi_{1}^{\prime }(x) & \chi_{2}^{\prime }(x)
\end{array}
\right) 
\]
and the resolvent matrix is:

\[
R(x,y)=V(y)V^{-1}(x) 
\]
It is the matrix linking the value of one solution at point $x$ to its value
at point $y$. The monodromy matrix is finally defined as: 
\[
{\bf T=R(}0,1)=\left( 
\begin{array}{ll}
\chi_{1}(1) & \chi_{2}(1) \\ 
\chi_{1}^{\prime }(1) & \chi_{2}^{\prime }(1)
\end{array}
\right) 
\]
This matrix characterizes a layer as it allows the writing of the matching
between the boundary conditions. Indeed, taking into account the propagation
conditions (\ref{propag}), we derive the following relation:

\begin{equation}
{\bf T}^{N}{%
{1+r_{N}  \choose i\beta _{0}(1-r_{N})}%
}=t_{N}{%
{1  \choose i\beta _{0}}%
}  \label{raccord}
\end{equation}

which permits to obtain both values of $r_{N}$ and $t_{N}$. When dealing
with a wavepacket, we write the reflected field under the form:

\begin{equation}
U_{N}^{d}(x,y,t)=\int_{{\bf P}}r_{N}(k,\theta )\exp (ik(-\cos \theta x+\sin
\theta y))\exp (-i\omega t)d\mu \; .  \label{reflect}
\end{equation}
{\bf Our aim is now to study the limit behavior of }$r_{N}${\bf \ as }$N$%
{\bf \ tends to infinity.}

\subsection{Some properties of matrix ${\bf T}$. Connection with Bloch wave
theory}

The determinant of matrix ${\bf T}$ is the value of the wronskian of
solutions $w_{1}$ and $w_{2}$ at point $1$. As in that case the wronskian is
constant, it is equal to $1$. Consequently, ${\bf T}$ is a unimodular matrix
and its characteristic polynomial is $X^{2}-tr\left( {\bf T}\right) X+1$.
Therefore the eigenvalues of ${\bf T}$ are real if and only if $\frac{1}{2}%
\left| tr({\bf T)}\right| >1$. This suggests a splitting of the set of
parameters:

\begin{eqnarray*}
{\bf G}& =\left\{ \left( k,\theta \right) \in {\bf P}, \frac{1}{2}\left| tr(%
{\bf T)}\right| >1\right\} \\
{\bf B}& =\left\{ \left( k,\theta \right) \in {\bf P},\frac{1}{2}\left| tr(%
{\bf T)}\right| <1\right\} \\
{\bf \Delta }& =\left\{ \left( k,\theta \right) \in {\bf P},\frac{1}{2}%
\left| tr({\bf T)}\right| =1\right\}
\end{eqnarray*}
When $\left( k,\theta \right) \in {\bf B}$, the eigenvalues of ${\bf T} $
are conjugate complex numbers of modulus $1$, whereas when $\left( k,\theta
\right) \in {\bf G}${\bf ,} ${\bf T}$ has two real eigenvalues $\gamma
\left( k,\theta \right) ,{\displaystyle{\frac{1 }{\gamma \left( k,\theta
\right) }}}$ where, by convention, $\left| \gamma \left( k,\theta \right)
\right| <1$. If $\left( k,\theta \right) \in {\bf \Delta }$, then ${\bf T}$
has an eigenvalue $\gamma \left( k,\theta \right) $ of multiplicity $2$ with
either $\gamma \left( k,\theta \right) =1$ or $\gamma \left( k,\theta
\right) =-1$. We denote by ${\bf \Delta }_{0}$ the subset of ${\bf \Delta }$
where ${\bf T} $ or $-{\bf T}$ is the identity matrix.

Bloch wave theory is the convenient tool when dealing with propagation
equations with periodic coefficients. Given a periodic medium with period $Y$%
, it consists in searching solutions of Schr\"{o}dinger or wave equation
under the form $u_{\overrightarrow{k}}\left( \overrightarrow{x}\right) e^{i%
\overrightarrow{k}.\overrightarrow{x}}$, where $u_{\overrightarrow{k}}\left( 
\overrightarrow{x}\right) $ is a $Y$-periodic function and $\overrightarrow{k%
}$ belongs to the so-called first Brillouin zone $Y^{\prime }$. For
one-dimensional media, with period $Y=\left[ 0,1\right[ $, the theory is
quite simple, for in that case $Y^{\prime }=\left[ -\pi ,+\pi \right[ $.
Therefore, Bloch solutions write $v_{\phi }\left( x\right) =u_{\phi }\left(
x\right) e^{i\phi x}$, where the so-called Bloch frequency $\phi $ belongs
to $Y^{\prime }$, and $u_{\phi }\left( x+1\right) =u_{\phi }\left( x\right) $%
. Thus we have $v_{\phi }\left( x+1\right) =e^{i\phi }v_{\phi }\left(
x\right) $. From the definition of ${\bf T}$ this means that $e^{i\phi }$ is
an eigenvalue of ${\bf T}$, and the above remarks shows that $tr({\bf T}%
)=2\cos \phi $, which provides us with the dispersion relation of the medium
and leads us to define $\phi $ a function on ${\bf B}$ by: 
\begin{equation}
\phi \left( k,\theta \right) =\arccos \left( \frac{1}{2}tr({\bf T})\right)
\end{equation}
Obviously, for couples $\left( k,\theta \right) $ belonging to ${\bf B}$, it
is possible to define a Bloch frequency $\phi $, and therefore there exists
propagating waves in the medium: such couples $\left( k,\theta \right) $
thus define a conduction band. If $\left( k,\theta \right) $ belongs to $%
{\bf G}$ then there are only evanescent waves and $\left( k,\theta \right) $
belongs to a forbidden band.

\subsection{Explicit expression of the reflection coefficient}

We suppose that $\left( k,\theta \right) $ belongs to ${\bf P\backslash
\Delta }$. Denoting $\left( {\bf v},{\bf w}\right) $ a basis of eigenvectors
of ${\bf T}$ we write in the canonical basis of ${\Bbb {R}^{2}}$~: ${\bf v=}%
\left( v_{1},v_{2}\right) ,{\bf w=}\left( w_{1},w_{2}\right) .$ Eigenvector $%
{\bf v}$ (resp. ${\bf w}$) is associated to eigenvalue $\gamma \left(
k,\theta \right) $ (resp. $\gamma ^{-1}\left( k,\theta \right) $). It is of
course always possible to choose $\left( {\bf v},{\bf w}\right) $ such that $%
\det \left( {\bf v},{\bf w}\right) =1$. After tedious but easy calculations,
we get $r_{N}$ in closed form from (\ref{raccord}): 
\begin{eqnarray}
r_{N}\left( k,\theta \right) &=&\frac{\left( \gamma ^{2N}-1\right) f}{\gamma
^{2N}-g^{-1}f}  \label{reflec} \\
t_{N}\left( k,\theta \right) &=&\frac{\left( 1-g^{-1}f\right) \gamma ^{N}}{%
\gamma ^{2N}-g^{-1}f}
\end{eqnarray}
denoting $q(x_{1},x_{2})={\displaystyle{\frac{i\beta _{0}x_{2}-x_{1}}{i\beta
_{0}x_{2}+x_{1}}}}$\ , functions $f$ and $g$ are defined by 
\begin{equation}
\begin{array}{ll}
\hbox{if }\left( k,\theta \right) \in {\bf G} & g\left( k,\theta \right)
=q\left( {\bf v}\right) ,f\left( k,\theta \right) =q\left( {\bf w}\right) \\ 
\hbox{if }\left( k,\theta \right) \in {\bf B} & \left\{ 
\begin{array}{l}
g\left( k,\theta \right) =q\left( {\bf v}\right) ,f\left( k,\theta \right)
=q\left( {\bf w}\right) \hbox{ if }\left| q\left( {\bf v}\right) \right|
<\left| q\left( {\bf w}\right) \right| \\ 
g\left( k,\theta \right) =q\left( {\bf w}\right) ,f\left( k,\theta \right)
=q\left( {\bf v}\right) \hbox{ if }\left| q\left( {\bf w}\right) \right|
<\left| q\left( {\bf v}\right) \right|
\end{array}
\right.
\end{array}
\end{equation}
\newline
{\bf Remark: }We have $\left| g\right| <\left| f\right| $ and in a
conduction band $f=\overline{g}^{-1}$ so that we always have $\left| g\left(
k,\theta \right) \right| \leq 1$.

Let us denote finally denote: 
\begin{equation}
\begin{array}{c}
H_{r}\left( z\right) =f{\displaystyle{\frac{z^{2}-1}{z^{2}-g^{-1}f}}} \\ 
H_{t}=\left( 1-g^{-1}f\right) {\displaystyle{\frac{z}{z^{2}-g^{-1}f}}}
\end{array}
,
\end{equation}
We immediately see that the reflection and transmission coefficients are
obtained through $H_{r}$ and $H_{t}$ by: 
\[
r_{N}\left( k,\theta \right) =H_{r}\left( \gamma ^{N}\right) ,t_{N}\left(
k,\theta \right) =H_{t}\left( \gamma ^{N}\right) . 
\]

\section{Asymptotic analysis of the reflection coefficient}

The reflection coefficient defines a sequence of points belonging to the
closed unit disc ${\Bbb {D}}$ of the complex plane. In order to have a clear
understanding of the behavior of $\left\{ r_{N},t_{N}\right\} $, we
interpret this sequence as a discrete dynamical system. As $N$ increases, we
want to study how the orbits of this system spread over ${\Bbb {D}}$.

When $\left( k,\theta \right) $ belongs to ${\bf G}\cup \left( {\bf \Delta
\setminus \Delta }_{0}\right) $ the behavior of the dynamical system is
trivial as $\left\{ r_{N},t_{N}\right\} $ admits one cluster point, situated
on ${\Bbb {U=}\left\{ z,\left| z\right| =1\right\} }$. Indeed, in that case $%
\gamma $ belongs to ${\Bbb {R \setminus}\left\{ -1,1\right\} }$. Recalling
that by convention $\left| \gamma \right| <1$, we see that $H_r\left( \gamma
^{N}\right) $ tends to $g\left( k,\theta \right) $. As the eigenvectors of $%
{\bf T}$ are real, we conclude that $g\left( k,\theta \right) $ belongs to $%
{\Bbb {U}}$. The second easy case is for $\left( k,\theta \right) $
belonging to ${\bf \Delta }_{0}$, indeed $\left\{ r_{N},t_{N}\right\} $ is
constant and equal to $0$ whatever $N$.

These two cases hand the gaps as well as the edges of the gaps. The case of
the conduction bands, i.e. for $\left( k,\theta \right) $ belonging to ${\bf %
B}$ is much more complicated and interesting. In the following we will skip
the mathematical rigor and stress on the physical meaning of the results.
The interested reader will find a complete and rigorous mathematical
discussion in \cite{semi-inf}.

Dealing with a couple $\left( k,\theta \right) $ belonging to ${\bf B}$, the
eigenvalues of ${\bf T}$ now belong to ${\Bbb {U}}$ and we may write $\gamma
=e^{i\phi }$. Obviously, $r_{N}=H_r\left( e^{iN\phi }\right) $ has no
pointwise limit as $N$ tends to infinity. From a geometrical point of view,
it is easy to show that the image of ${\Bbb {U}}$ through $H$ is a circle $%
{\bf V}\left( k,\theta \right) $ passing through the origin and whose
cartesian equation writes:

\begin{equation}
\left( x-\Re {\it e}\{z_{0}\}\right) ^{2}+\left( y-\Im {\it m}%
\{z_{0}\}\right) ^{2}=\left| z_{0}\right| ^{2}\quad 
\hbox{, with $z_0 = (
f^{-1} + g^{-1})^{-1}$ }
\end{equation}
Therefore, $\left\{ r_{N},t_{N}\right\} $ describes a set of point on ${\bf V%
}\left( k,\theta \right) $. As $\left\{ r_{N},t_{N}\right\} $ does not
converge pointwise, we turn to another notion of convergence, in some
average meaning. An easy computation in the case $\left| g(k,\theta )\right|
<\left| f(k,\theta )\right| $, shows that 
\begin{eqnarray}
r_{N}\left( k,\theta \right) &=&g+g\sum_{k=1}^{+\infty }\gamma ^{2Nk}\left[
\left( gf^{-1}\right) ^{k}-\left( gf^{-1}\right) ^{k-1}\right] \\
t_{N}\left( k,\theta \right) &=&(1-gf^{-1})\gamma ^{N}\sum_{k=0}^{+\infty
}\gamma ^{2Nk}\left( gf^{-1}\right) ^{k}
\end{eqnarray}
\newline
then the reflected and transmitted fields write 
\begin{eqnarray*}
U_{N}^{d}(x,y,t) &=&\int_{{\bf P}}g\left( k,\theta \right) \exp ^{i\left( 
{\bf k}\cdot {\bf r}-\omega t\right) }d\mu \\
&+&\sum_{k=1}^{+\infty }\int_{{\bf P}}\gamma ^{2Nk}g\left( k,\theta \right) %
\left[ \left( gf^{-1}\right) ^{k}-\left( gf^{-1}\right) ^{k-1}\right] \exp
^{i({\bf k}\cdot {\bf r}-\omega t)}d\mu \\
U_{N}^{t}(x,y,t) &=&\sum_{k=1}^{+\infty }\int_{{\bf P}}\gamma ^{\left(
2k+1\right) N}(1-gf^{-1})\left( gf^{-1}\right) ^{k}\exp ^{i({\bf k}\cdot 
{\bf r}-\omega t)}d\mu
\end{eqnarray*}

{\bf Definition 1: }{\it We say that a sequence of functions }$\psi
_{N}\left( k,\theta \right) ${\it \ converges weakly towards }$\psi _{\infty
}\left( k,\theta \right) ${\it \ if }$%
\mathrel{\mathop{\lim
}\limits_{N\longrightarrow +\infty }}\int_{{\bf B}}\psi _{N}\left( k,\theta
\right) \varphi \left( k,\theta \right) d\mu =\int_{{\bf B}}\psi _{\infty
}\left( k,\theta \right) \varphi \left( k,\theta \right) d\mu ${\it \ for
every }$\varphi ${\it \ belonging to }$L^{1}\left( {\bf B,}\mu \right) ${\it %
.}

We want to pass to the limit $N\rightarrow +\infty $ in the preceding
expressions. What we expect is some averaging over the set $Y^{\prime }$.
Clearly, the limit behavior relies on the properties of $\mu $ and $\phi $.
Let us define a convenient class of measures. Denoting $C_{\#}\left(
Y^{\prime }\right) $ the space of continuous $Y^{\prime }$-periodic
functions, we put:

{\bf Definition 2}: {\it A measure }$\mu ${\it \ is said admissible, if }$%
\exp \left( iN\phi \left( k,\theta \right) \right) ${\it \ tends weakly to} $%
0$.

Of course, this looks like an {\it ad hoc} property as it allows to get
directly the limits of interest, but indeed this is a correct way of
handling the problem, as it can be shown that the measures of interest for
our problem, i.e. that of physical significance, will prove to be
admissible. From the above expression, we conclude that 
\begin{eqnarray*}
U_{N}^{d}(x,y,t) &\rightarrow &\int_{{\bf P}}g\left( k,\theta \right) \exp
^{i\left( k.r-\omega t\right) }d\mu \\
U_{N}^{t}(x,y,t) &\rightarrow &0
\end{eqnarray*}

We can now conclude by collecting the above results.

{\bf Proposition 1}: {\it As }$N${\it \ tends to infinity, }$r_{N}${\it \
converges }weakly{\it \ towards }$r_{\infty }\left( k,\theta \right)
=g\left( k,\theta \right) ${\it , }$t_{N}$ tends weakly to $0$.

\section{Diffraction of a wavepacket by a semi-infinite medium}

Now, let us choose the incident field as a wavepacket of the form (\ref
{incident}) where $\mu$ is of one of the following forms:

\begin{eqnarray*}
{\bf I}& :\mu =p(k,\theta )dk\otimes d\theta ,p\in L^{1}({\bf P},dk\otimes
d\theta ) \\
{\bf II}& :\mu =p(\theta )\delta _{k}\otimes d\theta ,p\in L^{1}(\left] -%
\frac{\pi }{2},+\frac{\pi }{2}\right[ ,d\theta ) \\
{\bf III}& :\mu =p(k)dk\otimes \delta _{\theta },p\in L^{1}({\bf R}%
^{+},dk\otimes \delta _{\theta })
\end{eqnarray*}
These measures define the most commonly used incident fields. Indeed
measures of type ${\bf I}$ correspond to a general wavepacket, measures of
type ${\bf II}$ to a monochromatic beam, and measures of type ${\bf III}$ to
a temporal pulse. In order to apply the above results we have the following
fundamental result:

{\bf Proposition 2}: {\it Measures of type }${\bf I,II,III}${\it \ are
admissible.}

We can conclude that, for these measures $\mu ,$ the diffracted field $%
U_{N}^{d}\left( x,y,t\right) $ corresponding to $N$ layers converges
uniformly towards $U_{\infty }^{d}(x,y,t)$ given by:

\begin{equation}
U_{\infty }^{d}(x,y,t)=\int_{{\bf P}}r_{\infty }(k,\theta )\exp (ik(\sin
\theta x+\cos \theta y))\exp (-i\omega t)d\mu
\end{equation}
That way, we have obtained a rigorous formulation for the diffraction of a
wavepacket by a semi-infinite medium.

\section{Reflection of a monochromatic beam by a lossless slab of infinite
thickness}

A very special and interesting case is the case of a simple slab of optical
index $\nu $: the reflection coefficient ($r_{h}$) is therefore well known : 
\begin{equation}
r_{h}=r_{dio}(\frac{1-\exp ^{2i\beta h}}{1-r_{dio}^{2}\exp ^{2i\beta h}})
\label{reflection}
\end{equation}
where $r_{dio}$ is the reflection coefficient for the diopter ($r_{dio}={%
\displaystyle{\frac{\beta _{0}-\beta }{\beta _{0}+\beta }}}$). When the slab
is filled with a lossy material $N$ is therefore element of ${\Bbb {C}-\{{R}%
,i{R}\}}$, and in that case: 
\begin{equation}
\forall (k_{0},\theta )\in {\bf P},r_{h}\longrightarrow r_{dio}
\end{equation}
On the contrary, when we are dealing with lossless materials, $\nu $ is
real. In that case, $r_{h}$ has an oscillating behavior and does not
converge (and {\em a fortiori} does not converge to $r_{dio}$). However, if
we consider a limited monochromatic incident beam described by a function $%
u_{inc}(x,y)$ : 
\begin{equation}
u_{inc}(x,y)=\int_{-k_{0}}^{k_{0}}p(\alpha )e^{i(\alpha x+\beta (\alpha
)y)}\,d\alpha
\end{equation}
where $p(\alpha )\in L^{1}(]-k_{0},k_{0}[,d\alpha )$ and characterizes the
shape of the incident beam. We are therefore in the case where the measure $%
\mu $ is of the form ${\bf II}$ (cf. paragraph V) i.e. : 
\begin{equation}
\mu =p(\theta )\delta _{k}\otimes d\theta
\end{equation}
with $p\in L^{1}(]-\pi /2,\pi /2[,d\theta )$. In these conditions the
diffracted (reflected) field $u_{h}(x,y)$ corresponding to a slab of finite
thickness $h$ follows as such : 
\begin{equation}
u_{h}(x,y)=\int_{-k_{0}}^{k_{0}}r_{h}(\alpha )p(\alpha )e^{i(-\alpha x+\beta
(\alpha )y)}\,d\alpha \;
\end{equation}
and the diffracted field $u_{dio}(x,y)$ corresponding to the diopter : 
\begin{equation}
u_{dio}(x,y)=\int_{-k_{0}}^{k_{0}}r_{dio}h(\alpha )p(\alpha )e^{i(-\alpha
x+\beta (\alpha )y)}\,d\alpha \;
\end{equation}
and we have the fundamental result : 
\begin{equation}
\forall (x,y)\in {\bf R}\times {\bf R}_{+}, u_{h}(x,y)\longrightarrow
u_{\infty }(x,y)=\int_{-k_{0}}^{k_{0}}r_{\infty }(\alpha )p(\alpha
)e^{i(-\alpha x+\beta (\alpha )y)}\,d\alpha \;
\end{equation}
It is very easy to calculate the reflection coefficient $r_{\infty }$ in
that case. The monodromy matrix is: 
\begin{equation}
{\bf T}=\left( 
\begin{array}{ll}
\cos \left( \beta h\right) & \frac{1}{\beta }\sin \left( \beta h\right) \\ 
-\beta \sin \left( \beta h\right) & \cos \left( \beta h\right)
\end{array}
\right)
\end{equation}
so that $tr\left( {\bf T}\right) =2\cos \beta h,$ therefore every $\left(
k,\theta \right) $ belongs to ${\bf B\cup \Delta }_{0}$ and $\phi =\arccos
(1/2Tr({\bf T}))=\beta h$. After elementary calculations we find : 
\begin{equation}
g=\frac{\beta _{0}-\beta }{\beta _{0}+\beta }=r_{dio}\quad \quad f=(\frac{%
\beta _{0}-\beta }{\beta _{0}+\beta })^{-1}
\end{equation}
Taking everything into account, we find the weak convergence $r_{\infty
}=r_{dio}$ and therefore $u_{h}\longrightarrow u_{dio}$.

\section{Electromagnetic Perot Fabry}

In numerous experiments in Optics it appears that the light concentrates
round particular areas of the overall space, namely the rays. In this
paragraph, we establish a link between the amplitude associated with a ray
and the amplitude associated with the electric field.

\subsection{Simple Perot Fabry}

It is well known that many reflected rays appear when dealing with the
reflection coefficient of a monochromatic beam by a ``simple'' Perot-Fabry,
we are in the same case as in the precedent paragraph, for wich the
reflected field corresponding to the finite thickness slab $u_{h}(x,y)$%
writes as follows: 
\begin{equation}
u_{h}(x,y)=\int_{-k_{0}}^{k_{0}}r_{h}(\alpha )p(\alpha )e^{i(-\alpha x+\beta
(\alpha )y)}\,d\alpha  \label{uhxy}
\end{equation}
>From equation (\ref{reflection}), the reflection coefficient $r_{h}$can be
expressed as a series in the following manner:

\begin{equation}
r_{h}=r_{dio}+\sum_{l=1}^{+\infty }(r_{dio}^{2l+1}-r_{dio}^{2l-1})e^{2i\beta
hl}
\end{equation}
Consequently, $u_{h}$can also be expressed as a series (the exponential
decreasing of $r_{dio}$ allows us to reverse the signs $\sum $and $\int $):

\begin{equation}
u_{h}(x,y)=u_{dio}(x,y)+\sum_{l=1}^{+\infty }u_{h,l}(x,y)  \label{uhser}
\end{equation}
with

\begin{equation}
u_{h,l}(x,y)=\int_{-k_{0}}^{k_{0}}r_{dio}^{2l-1}(r_{dio}^{2}-1)e^{2i\beta
(\alpha )hl}p(\alpha )e^{i\left( -\alpha x+\beta (\alpha )y\right) }\,d\alpha
\end{equation}

Finally, introducing the two transmission coefficients $t_{12}={\displaystyle%
{\frac{2\beta _{0} }{\beta _{0}+\beta }}}$ and $t_{21}={\displaystyle{\frac{%
2\beta }{\beta _{0}+\beta }}}$, we find an expression of $u_{h,l}$ that the
opticist can interpret at first glance:

\begin{equation}
\left\{ 
\begin{array}{l}
u_{h,l}(x,y)=-\int_{-k_{0}}^{k_{0}}r_{dio}^{2l-1}t_{12}t_{21}e^{2i\beta
(\alpha )hl}p(\alpha )e^{i(-\alpha x+\beta (\alpha )y)}\,d\alpha ,l\geq 1 \\ 
u_{h,0}(x,y)=u_{dio}(x,y)
\end{array}
\right.
\end{equation}

Each function $u_{h,l}$ , $l\in \left\{ 0,\cdots ,+\infty \right\} $ is
interpreted as the complex amplitude associated with a ray labeled by $l$
(cf. figure 1):

\begin{itemize}
\item  the term $r_{dio}^{2l-1}t_{12}t_{12}$ corresponds to the amplitude
associated with a ray $l$.

\item  the term $e^{2i\beta (\alpha )hl}$ corresponds to a term of phase
which expresses the delay of the reflected ray $l$ with respect to the first
reflected ray ($l=0$)

\item  the minus sign before the integral may appear as suspicious. However,
for the opticist, the interpretation of this sign is easy. The reflection of
all rays are of the same nature (reflection of a ray from a medium of index $%
\nu $ on a medium of index $1$) except for the first ray (reflection of a
ray from a medium of index $1$ on a medium of index $\nu $), which implies a
change of sign in the reflection coefficient.
\end{itemize}

Besides, expression (\ref{uhser}) calls for two remarks :

\begin{enumerate}
\item  the optical interpretation of functions $u_{h,l}$ is all the more
clear as the following conditions are better fulfilled: a) the supports of
the rays $l$ are actually separated, i.e. the function $p$ has a narrow
support , b) the incident field is sufficiently slanted and c) the depth of
the slab is sufficiently large with respect to the wavelength.

\item  in the precedent paragraph, we have shown that $u_{h}$ tends to $%
u_{dio}$ as $h$ tends to infinity, consequently the series in equation (\ref
{uhser}) has to tend to zero when $h$ tends to infinity. Therefore, each
function $u_{h,l}$ behaves as a corrector for the diffracted field (each ray
except the first one does vanish when the depth of the slab tends to
infinity).
\end{enumerate}

\subsection{Generalized Perot Fabry}

We consider now a one dimensional structure made of $N$ identical layers.
Writing $u_{N}(x,y)$ the field diffracted by this structure, we have:

\begin{equation}
u_{N}(x,y)=\int_{-k_{0}}^{k_{0}}r_{N}(\alpha )p(\alpha )e^{i(-\alpha x+\beta
(\alpha )y)}\,d\alpha
\end{equation}
Using the expression of $r_{N}$ in (\ref{reflect}) and the same techniques
used above, we find :

\begin{equation}
u_{N}(x,y)=u_{\infty }(x,y)+\sum_{l=1}^{+\infty }u_{N,l}(x,y)
\end{equation}
with

\begin{equation}
u_{\infty }(x,y)=\int_{-k_{0}}^{k_{0}}r_{\infty }(\alpha )p(\alpha
)e^{i(-\alpha x+\beta (\alpha )y)}\,d\alpha
\end{equation}
and

\begin{equation}
u_{N,l}(x,y)=\int_{-k_{0}}^{k_{0}}\gamma ^{2Nl}g\left( k,\theta \right) 
\left[ \left( gf^{-1}\right) ^{l}-\left( gf^{-1}\right) ^{l-1}\right]
p(\alpha )e^{i(-\alpha x+\beta (\alpha )y)}\,d\alpha
\end{equation}
The same comments as in the preceding section can be made.

\subsection{Speed of convergence : numerical experiments}

Up to now a special attention has been drawn about theoretical aspects of
electromagnetic diffraction. More precisely, we have proved that, in any
case encountered in physics, the function $u_{N}(x,y,t)$ which represents
the diffracted field converges to a function $u_{\infty }(x,y,t)$ which one
can easily calculate. But it remains to be seen how the function $%
u_{N}(x,y,t)$ converges to its limit. For instance, in harmonic regime with
a pulsatance $\omega $, it is of practical prime importance to know the
number of layers $N_{\eta }$ from which one can replace, with a given
precision $\eta $, $u_{N}(x,y,\omega )$ by $u_{\infty }(x,y,\omega )$. From
a theoretical point of view, it is very difficult to answer such questions
even if , in some respects, we outlined an answer in the precedent
paragraph. We are thus ``doomed'' to make numerical experiments and to leave
the general insights of the Theory. Of course, in this paragraph, we do not
aspire to the exhaustiveness and we only aim at giving some rough estimates.

In what follows, we are only dealing with a single Perot-Fabry made of
silice for which the refractive index is $\nu =1.47$ illuminated by a
monochromatic gaussian beam solely characterized by the wavelength ($\lambda
={\displaystyle{\frac{2\pi c }{\omega }}}$ ), the size $\Delta x_{w}$ and
the height $x_{w}$ of the waist and the mean incidence angle $\theta $ (cf.
figure \ref{fig1}). For such a beam the function $p(\alpha )$ introduced
above writes as follows:

\begin{equation}
p(\alpha )=\frac{1}{2\pi }\exp \left( -\left( \frac{\alpha -\alpha _{0}}{%
2\Delta x_{w}}\right) ^{2}\right) \exp \left( -i(\alpha _{0}x_{w}-\beta
_{0}y_{w})\right)
\end{equation}
where $\alpha _{0}={\displaystyle{\frac{2\pi }{\lambda }}}\sin \left( \theta
\right) $ and $y_{w}=-\tan \left( \theta \right) x_{w}$. In our numerical
experiments, we have taken the following values : $\theta ={\displaystyle{%
\frac{\pi }{4}}}$, ${\displaystyle{\frac{x_{w} }{\lambda }}}=200$ and ${%
\displaystyle{\frac{H }{\lambda }}}=200$, where $H$ is the ``height of
observation'' (cf. figure \ref{fig1}). For these values, we have drawn, in
the same figure $\left| u_{h}(H,y)\right| ^{2}$, $\left| u_{\infty
}(H,y)\right| ^{2}$and $\left| u_{h,1}(H,y)\right| ^{2}$as functions of $y$
(shifted of $\frac{y_{0}}{\lambda }$) for the following normalized depths (${%
\displaystyle{\frac{h }{\lambda }}}=2,20,200,2000,5000$) and for two
normalized waists (${\displaystyle{\frac{\Delta x_{w} }{\lambda }}}=5,50$)
(cf. figures \ref{fig3} to \ref{fig7} for a waist of $5$, and figures $8$ to 
$12$ for a waist of $50$). Geometrical Optics predicts the following
locations of the maxima (cf. figure \ref{fig2}):

\begin{center}
{\ 
\begin{tabular}{|c|c|c|c|c|c|}
\hline
$\frac{h}{\lambda }$ & $2$ & $20$ & $200$ & $2000$ & $5000$ \\ \hline\hline
$\frac{y_{0}}{\lambda }$ & $100$ & $100$ & $100$ & $100$ & $100$ \\ \hline
$\frac{y_{1}}{\lambda }$ & $102.2$ & $121.9$ & $319.5$ & $2295$ & $5587$ \\ 
\hline
\end{tabular}
}
\end{center}

\vspace{0.3cm}

Finally, we have drawn in figure \ref{fig13} (resp. fig. \ref{fig14}) the
total field map $\left| u_{h}(x,y)+u_{inc}(x,y)\right|$ for ${\displaystyle{%
\frac{\Delta x_{w} }{\lambda }}}=5$ (resp. for${\displaystyle{\frac{\Delta
x_{w} }{\lambda }}}=50$) and for depth ${\displaystyle{\frac{h }{\lambda }}}%
=200$.

In this example, we see that a width of about 1000 wavelengths (for $\lambda
=0.5\mu m$\- this means a $0.5mm$ width for the substrate) is necessary to
obtain the rays described in classical optics. In practical experiments, the
width of the substrate is usually of the order of one millimeter or more,
and therefore the Fresnel coefficients are indeed measured.

\section{Conclusion}

What is classical Optics (or coherent Optics)? Since Maxwell, at the end of
the nineteenth century, the answer seems to be easy: classical optics is the
study of the diffraction of an electromagnetic field by a body whose size is
very large compared with the wavelength (mean wavelength). That means that
we have to consider classical optics as a limit for small wavelengths of
electromagnetic optics. Unlike the {\it doxa, }we consider that this limit
is far from being clear. For instance, it is well known that in
electromagnetic optics, the diffracted field is very sensitive to the
polarization of the incident field. On the contrary, for small wavelengths,
in a lot of applications, the diffracted field is independent of the
polarization. In many cases, this remarkable property remains mysterious. In
this paper, we only dealt with one dimensional problem and we demonstrated
that the limit analysis (the path from the electromagnetic problem to the
optical one) is generally ill posed for a plane wave (Fortunately, this case
does not exist!) but is well posed for the cases encountered in physics.
Besides, the limit analysis that we have given above does not take into
account the roughness of the layers. For the sake of simplicity, we did not
think fit to describe in a realistic way the process of measuring the field:
this would have lead us to use a spatial convolution process. These remarks
are not at all limitations of our study. On the contrary, both these
phenomena would have improved the convergence of the involved sequences.
Nevertheless we do think that this Limit Analysis is far more fundamental
for a good understanding of classical optics than the usual
explanations.Take the case of spectacles, which are usually intended to
improve the vision. Can one be satisfied by such an explanation as that
involving absorption? The challenge is now to extend these results to the
bidimensional or even tridimensional case, for which the very mathematical
formulation itself is far from being clear.

{\bf Acknowledgment}

We wish to thank Pr. Guy Bouchitt\'{e} for constructive remarks on way to
improve this paper and for enlightning many mathematical points. We are also
gratefully indebted to Mrs. Margaret Gaillard for careful reading of the
manuscript.

\newpage

{\bf Figures Captions}

Figure 1: Experimental device, position of the waist and height of
observation.

Figure 2: Location of the maxima foreseen by geometrical optics.

Figure 3: Wavelength ($\lambda =1$) , $n=1.47$, size of the waist ($\Delta
x_{w}=5$), position of the waist ($x_{w}=200,y_{w}=-200$), mean incidence
angle ($\theta =\pi /4$), height of observation ($H=100$), depth of layer $%
h=2$.

Figure 4: Wavelength ($\lambda =1$) , $n=1.47$, size of the waist ($\Delta
x_{w}=5$), position of the waist ($x_{w}=200,y_{w}=-200$), mean incidence
angle ($\theta =\pi /4$), height of observation ($H=100$), depth of layer $%
h=20$.

Figure 5: Wavelength ($\lambda =1$) , $n=1.47$, size of the waist ($\Delta
x_{w}=5$), position of the waist ($x_{w}=200,y_{w}=-200$), mean incidence
angle ($\theta =\pi /4$), height of observation ($H=100$), depth of layer $%
h=200$.

Figure 6: Wavelength ($\lambda =1$) , $n=1.47$, size of the waist ($\Delta
x_{w}=5$), position of the waist ($x_{w}=200,y_{w}=-200$), mean incidence
angle ($\theta =\pi /4$), height of observation ($H=100$), depth of layer $%
h=2000$.

Figure 7: Wavelength ($\lambda =1$) , $n=1.47$, size of the waist ($\Delta
x_{w}=5$), position of the waist ($x_{w}=200,y_{w}=-200$), mean incidence
angle ($\theta =\pi /4$), height of observation ($H=100$), depth of layer $%
h=5000$.

Figure 8: Wavelength ($\lambda =1$) , $n=1.47$, size of the waist ($\Delta
x_{w}=50$), position of the waist ($x_{w}=200,y_{w}=-200$), mean incidence
angle ($\theta =\pi /4$), height of observation ($H=100$), depth of layer $%
h=2$.

Figure 9: Wavelength ($\lambda =1$) , $n=1.47$, size of the waist ($\Delta
x_{w}=50$), position of the waist ($x_{w}=200,y_{w}=-200$), mean incidence
angle ($\theta =\pi /4$), height of observation ($H=100$), depth of layer $%
h=20$.

Figure 10: Wavelength ($\lambda =1$) , $n=1.47$, size of the waist ($\Delta
x_{w}=50$), position of the waist ($x_{w}=200,y_{w}=-200$), mean incidence
angle ($\theta =\pi /4$), height of observation ($H=100$), depth of layer $%
h=200$.

Figure 11: Wavelength ($\lambda =1$) , $n=1.47$, size of the waist ($\Delta
x_{w}=50$), position of the waist ($x_{w}=200,y_{w}=-200$), mean incidence
angle ($\theta =\pi /4$), height of observation ($H=100$), depth of layer $%
h=2000$.

Figure 12: Wavelength ($\lambda =1$) , $n=1.47$, size of the waist ($\Delta
x_{w}=50$), position of the waist ($x_{w}=200,y_{w}=-200$), mean incidence
angle ($\theta =\pi /4$), height of observation ($H=100$), depth of layer $%
h=5000$.

Figure 13: The total field map $\left| u_{h}(x,y)+u_{inc}(x,y)\right| ^{2}$
with wavelength ($\lambda =1$) , $n=1.47$, size of the waist ($\Delta
x_{w}=50$), position of the waist ($x_{w}=200,y_{w}=-200$), mean incidence
angle ($\theta =\pi /4$), depth of layer $h=200$.

\end{document}